\documentclass[aps,superscriptaddress,letterpaper,]{revtex4}

\RequirePackage{amsfonts,amsmath,amssymb,bm}

\begin{document}


\title{Errata: Lattice-corrected strain-induced vector potentials in
graphene \par [Phys. Rev. B 85, 115432 (2012)]}

\author{Alexander L Kitt}
\affiliation{Department of Physics, Boston University, 590
Commonwealth Ave, Boston, Massachusetts 02215, USA}

\author{Vitor M. Pereira}
\affiliation{Graphene Research Centre and Department of Physics,
National University of Singapore, 2 Science Drive 3, Sinagpore 117542
}

\author{Anna K Swan}
\affiliation{Department of Electrical and Computer Engineering, Boston
University, 8 St Mary's St, Boston, Massachusetts 02215, USA }
\affiliation{Department of Physics, Boston University, 590
Commonwealth Ave, Boston, Massachusetts 02215, USA}
\affiliation{Photonics Center, Boston University, 8 St Mary's St,
Boston, Massachusetts 02215, USA}

\author{Bennett B Goldberg}
\affiliation{Department of Physics, Boston University, 590
Commonwealth Ave, Boston, Massachusetts 02215, USA}
\affiliation{Center for Nanoscience, Boston University, 8 St Mary's
St, Boston, Massachusetts 02215, USA}
\affiliation{Photonics Center, Boston University, 8 St Mary's St,
Boston, Massachusetts 02215, USA}

\date{\today}

\maketitle


\setlength{\parindent}{0em}
\setlength{\parskip}{1em}

We have discovered that our form for the strained positions of the carbon atoms in the graphene lattice was incomplete. In this \emph{errata} we show how the complete treatment changes our conclusions. In particular, the second of our two conclusions below is not true:
\begin{enumerate}
  \item To correctly describe the shift in the positions of the Dirac points from the reference (flat, undeformed) state to the strained state in terms of a strain-induced vector potential, the corrections arising from the deformation of the lattice ($\vec{A}_\text{latt}=\vec{A}_{\bm{K}_i}-\vec{A}_p$, where $\vec{A}_{\bm{K}_i}$ and $\vec{A}_p$ are defined in Eqs.~4 of the original paper) should be taken into consideration in addition to the changes in the nearest-neighbor hoppings (e.g., Eqs.~4 or Figs.~2).
  \item A non-uniform (yet smooth, so that the effective mass approximation is still meaningful) strain distribution endows $\vec{A}_\text{latt}$ with a position dependence which leads to a correction to the pseudomagnetic field: $\vec{B}_{\bm{K}_i} =\nabla\times\vec{A}_{\bm{K}_i}$. Fig.~3 showed the effect of this correction.
\end{enumerate}

The second conclusion is incorrect because, although the corrections $\vec{A}_\text{latt}$ are finite and, in general, have a position dependence, their rotational is identically zero and, thus, so is their contribution to the pseudomagnetic field. This has been pointed out recently by de Juan \emph{et al.}~\cite{deJuan:2012}. Below we elaborate on that, and on the reason why Fig.~3 apparently shows a non-zero $\vec{B}_{\bm{K}_i}$, when it should have been zero by construction. We trust the detail will benefit the reader.

Our original form for the strained nearest neighbor vectors was incomplete. Under the Cauchy-Born hypothesis, the position of the $i$-th atom in the deformed configuration, $\bm{R}_i$, is given with reference to the undeformed one, $\bm{r}_i$, in terms of the deformation field $\bm{u}(\bm{r})$:
\begin{equation}
  \bm{R}_{i}=\bm{r}_{i}+\bm{u}(\bm{r_i})
  \label{eq:deformation}.
\end{equation}
The electronic dispersion is affected by changes in the nearest-neighbor vectors $\bm{\delta}_{1,2,3}$ which, on account of \eqref{eq:deformation}, are given approximately by
\begin{equation}
\bm{\delta}_{i}^{\prime}(\bm{r})\simeq\bm{\delta}_{i}(\bm{r}
)+\left(\bm{\delta }_{i} \cdot\bm{\nabla}\right)\bm{u}(\bm{r})
= (\bm{1}+\bm{\nabla u})\cdot\bm{\delta }_{i}
\label{correct}
  ,
\end{equation}
where $\bm{\nabla}\bm{u}$ is the Jacobian of the displacement field known as the displacement gradient tensor:
\begin{equation*}
  [\bm{\nabla u}]_{ij} = u_{i,j}
  = \frac{u_{i,j}+u_{j,i}}{2} + \frac{u_{i,j}-u_{j,i}}{2}
  \equiv \tilde{\epsilon}_{ij} + \tilde{\omega}_{ij}
  \quad\longrightarrow\quad \bm{\nabla u} = \tilde{\bm{\epsilon}} +
\tilde{\bm{\omega}}
  ,
\end{equation*}
where $\tilde{\bm{\omega}}$ is the rotation tensor and $\tilde{\bm{\epsilon}}$ is the \emph{linear} strain tensor which is only one part of the full (Lagrange) strain tensor given by $\bm{\epsilon} = \tfrac{1}{2}(\bm{\nabla u} + \bm{\nabla u}^\top+\bm{\nabla u}^\top\bm{\nabla u}) = \tilde{\bm{\epsilon}} + \tfrac{1}{2}(\bm{\nabla u}^\top\bm{\nabla u})$.  Instead of using Eq.~\eqref{correct}, in our original paper we mistakenly took the strained nearest-neighbor vector to be $\bm{\delta}_{i}^{\prime}(\bm{r})\simeq (\bm{1}+\bm{\epsilon})\cdot\bm{\delta }_{i}$, a result that is only true in special cases.  In fact, even $\bm{\delta}_{i}^{\prime}(\bm{r})\simeq (\bm{1}+\tilde{\bm{\epsilon}})\cdot\bm{\delta }_{i}$ is only valid if the deformation does not involve local rotation ($\tilde{\bm{\omega}}=0$).

When the correct expansion for the strained position of the atoms is used, it becomes apparent that the lattice corrections cannot contribute to the pseudomagnetic field. Upon expansion around a corner of the BZ of the undeformed lattice, $\bm{K}$, the lattice corrections to the vector potential can be cast as \cite{deJuan:2012}
\begin{equation}
  \vec{A}_\text{latt}=\vec{A}_{\bm{K}}-\vec{A}_p \propto \bm{\nabla} (\bm{K}\cdot\bm{u})
  .
\end{equation}
Since the above is a total derivative, it cannot contribute to the pseudomagnetic field because $\nabla\times\nabla \phi \equiv 0$. This can be also verified by direct inspection of Eqs.~(4) of the original paper which remain valid in form if the replacement $\bm{\epsilon } \to \bm{ \nabla u }$ is made.

Having clarified and established the correct expansion of the nearest neighbor vector and the lack of an induced pseudomagnetic field a few comments on the results in the original paper are in order:
\begin{enumerate} \renewcommand{\theenumi}{\roman{enumi}}
  \item Lattice corrections beyond hopping clearly do not contribute to the pseudomagnetic field as originally claimed in the paper.
  \item The lattice corrections are still needed to correctly describe the shift in the positions of the Dirac points due to strain. Thus, when the position of the Dirac points is required in a global frame of reference (e.g. to describe momentum-sensitive electronic tunneling to/from strained graphene from/to another system, probe, or contact), Eqs.~4 should be used with the substitution $\bm{\epsilon } \to \bm{\nabla u}$:
	\begin{align*}
	\vec{A}_{\bm{K_1}}&=-\vec{A}_{\bm{K_1'}}=\frac{\phi_0}{2a} \left(
\begin{array}{c} \frac{4}{3\sqrt{3}}\bm{[\nabla u]}_{yx}\\ \frac{4}{3
\sqrt{3}} \bm{[\nabla u]}_{yy} \end{array} \right) +\vec{A}_p , \\ 
	\vec{A}_{\bm{K_2}}&=-\vec{A}_{\bm{K_2'}}=\frac{\phi_0}{2a} \left(
\begin{array}{c} \frac{2}{3}\bm{[\nabla u]}_{xx}-\frac{2\sqrt{3}}{9}
\bm{[\nabla u]}_{yx} \\ \frac{2}{3} \bm{[\nabla u]}_{xy}-\frac{2
\sqrt{3}}{9} \bm{[\nabla u]}_{yy} \end{array} \right)+\vec{A}_p  ,  \\
	\vec{A}_{\bm{K_3}}&=-\vec{A}_{\bm{K_3'}}=\frac{\phi_0}{2a} \left(
\begin{array}{c} -\frac{2}{3} \bm{[\nabla u]}_{xx}-\frac{2
\sqrt{3}}{9} \bm{[\nabla u]}_{yx} \\ -\frac{2}{3} \bm{[\nabla
u]}_{xy}-\frac{2 \sqrt{3}}{9} \bm{[\nabla u]}_{yy} \end{array}
\right)+\vec{A}_p  , \\
	\intertext{with}
	&\vec{A}_p=\frac{\phi_0}{2a} \left( \begin{array}{c} \frac{\beta }{\pi} \epsilon_{xy} \\ \frac{\beta }{2 \pi} (\epsilon_{xx}-\epsilon_{yy}) \end{array} \right).
	\end{align*}
  \item Unlike for the lattice corrections, the Lagrange strain tensor, $\bm{\epsilon}$, should still be used in the hopping contributions to the pseudo vector potential ($\vec{A}_p$) rather than $\bm{\nabla u}$ or $\tilde{\bm{\epsilon}}$. This is because, at the level of the tight-binding model considered here, $t_i$ depends only on the nearest-neighbor distance given to first order by: $|\bm{\delta}_{i}^{\prime}|\approx a+\frac{1}{a}\,\bm{\delta}_{i}\cdot\bm{\epsilon}\cdot\bm{\delta}_ {i}$.
  \item Eq.~(3) of the original paper,
\begin{equation*}
  H \simeq -\sum_{\vec{k},j} \bigl(
  t_0 + \delta t_j - it_0\vec{k}\cdot\epsilon\cdot\vec{\delta}_j
  \bigr) \, e^{-i\vec{k}\cdot\vec{\delta}_j} \,
  a_{\vec{k}}^{\dagger}b_{\vec{k}} + \text{H.c.}
  ,
  \label{eq:H-TB-strain}
\end{equation*}
is written as $\propto \bm{k}\cdot\bm{\epsilon}\cdot\bm{\delta}_i$, when, rigorously, it should have been $\propto\bm{k}\cdot\bm{\nabla u}\cdot\bm{\delta}_i$.
  \item Figs.~1 and 2 are correct.  For the deformations considered (plane, pure strain) the expressions in Eqs.~4 of the paper are correct as originally stated.
  \item Since lattice corrections cannot alter the pseudomagnetic field, Figs.~3(b) and 3(c) should be no different from Fig.~3(a).  This was inaccurate in the original paper because we expressed the deformation of the nearest-neighbor vectors in terms of the Lagrange strain tensor directly obtained from our finite element analysis rather than using the displacement gradient tensor. Extracting the plane projection of the displacement gradient from the simulation and replacing it in Eqs.~(4) yields a pseudomagnetic field with no contribution from lattice corrections. 
\end{enumerate}

We acknowledge discussions drawing our attention to these issues with M.~A.~H.~Vozmediano and G.~G.~Naumis.


\end{document}